\documentclass[RNAAS,longauthor]{aastex63}

\graphicspath{{./}{figures/}}

\begin{document}

\title{No Dispersed Single Radio Pulses Detected in Archival Parkes Pulsar Observations Targeting Supernova Remnants and Anomalous X-ray Pulsars}

\correspondingauthor{Fronefield Crawford}
\email{fcrawfor@fandm.edu}

\author[0000-0002-2578-0360]{Fronefield Crawford}
\affiliation{Department of Physics and Astronomy, Franklin and Marshall College, P.O. Box 3003, Lancaster, PA 17604, USA}

\begin{abstract}
Four supernova remnants and four anomalous X-ray pulsars were previously observed with the Parkes telescope in a campaign to detect pulsed radio emission from associated neutron stars. No signals were detected in the original searches of these data. I have reprocessed the data with the more recently developed HEIMDALL and FETCH software packages, which are optimized for single-pulse detection and classification. In this new analysis, no astrophysical pulses were detected having a signal-to-noise ratio above 7 from any of the targets at dispersion measures ranging from 0 to $10^{4}$ pc cm$^{-3}$. I include calculated fluence limits on single radio pulses from these targets.
\end{abstract}


\section{Introduction and Background} 

The Parkes 64-m radio telescope was used several decades ago to conduct deep search observations for pulsed radio signals from four supernova remnants (SNRs) that might harbor young radio pulsars and from four anomalous X-ray pulsars (AXPs). The Parkes multibeam receiver \citep{swb+96} and the 1-bit analog filterbank were used (see \citealt{mlc+01}). The targets were observed at a center frequency of 1374 MHz with a 288 MHz bandwidth split into 96 channels (see Table \ref{tbl-1}). Prior searches of these data for periodicities (from all of the targets) and for single pulses (from the AXPs) were unsuccessful \citep{cpk02,chk07}. Additional (unsuccessful) searches of separate observations of the four AXPs were also conducted by \citet{bri+06}. PSR J1747$-$2809 was subsequently discovered as a highly-scattered 52-ms pulsar associated with SNR G0.9+0.1 in 2 GHz observations from the Green Bank Telescope \citep{crg+09}. However, no detections were made in their (separate) 1.4 GHz Parkes observations of this source. Pulses at 1.4 GHz would be scattered by an amount comparable to the spin period \citep{crg+09}, which suggests that any sensitivity to single pulses from PSR J1474$-$2809 in the original Parkes observations is severely degraded. The fluence limit quoted Table \ref{tbl-1} does not take this into account and is an idealized limit for much narrower pulses. Three of the targets (SNR G327.1$-$1.1, SNR G347.5$-$0.5, and AX J1845.0$-$0258) are coincident with unassociated radio pulsars in the vicinity (PSRs B1550$-$54, J1713$-$3949, and J1844$-$0256, respectively). 

\section{Data Analysis and Results}

I used the HEIMDALL single-pulse detection package 
\citep{b12,bbb+12}\footnote{\url{https://sourceforge.net/projects/heimdall-astro}}
to search for dispersed radio pulses at dispersion measures (DMs) ranging from 0 to $10^{4}$ pc cm$^{-3}$, which significantly exceeds the maximum expected Galactic DM contribution along each target line of sight \citep{cl02,ymw17}. Boxcar matched filters were applied to each dedispersed time series to enhance sensitivity to a range of possible pulse widths. These filter windows ranged from 1 to 512 samples, which for most observations corresponded to an upper limit of 128 ms (see Table \ref{tbl-1}). After detection, the pulses were analyzed by the pulse classifier FETCH \citep{aab+20}\footnote{\url{https://github.com/devanshkv/fetch}}, which assigns a probability of being real to each detected pulse. The only pulses detected in the search above a signal-to-noise ratio of 7 that were classified as likely real by FETCH were four pulses from the known, bright pulsar PSR B1550$-$54, located near SNR G327.1$-$1.1 (but not associated with this SNR). All four pulses from PSR B1550$-$54 were clearly detected visually and were identified by FETCH as having a probability of being real greater than 50\%. The fluence limits from the non-detections of the targeted sources are listed in Table \ref{tbl-1}. These limits assume a pulse width of $W = 1$ ms and scale as $\sqrt{W}$ \citep{lk12}, and they include estimates of the sky temperature at the locations of the targets \citep{hss+82}. 

\begin{deluxetable}{lccccccc}
\tablecaption{Limits from Single-Pulse Searches of SNR and AXP Targets\label{tbl-1}}
\tablehead{
\colhead{Type} \vspace{-0.2cm} & \colhead{Target Name} & \colhead{RA (J2000)} & \colhead{Dec (J2000)} &  \colhead{Obs Date} & \colhead{$T_{int}$} & \colhead{Fluence Limit} & \colhead{Neutron Star} 
\\
\colhead{}   & \colhead{}    & \colhead{(hh:mm:ss)} & \colhead{(dd:mm:ss)}    & \colhead{(MJD)}   & \colhead{(s)} & \colhead{(Jy ms)} & \colhead{Association}
}
\startdata
SNR	 & G0.9+0.1$^{c}$          & 17:47:21 & $-$28:09:27 & 51377.44 & 14400 & 0.49 & PSR J1747$-$2809 \\
SNR	 & G266.2$-$1.2$^{b,c}$    & 08:52:00 & $-$46:22:00 & 51377.92 & 16800 & 0.31 & \ldots \\
AXP	 & 1E 1048.1$-$5937        & 10:50:09 & $-$59:53:20 & 51379.00 & 16800 & 0.27 & \ldots \\
AXP	 & 1RXS J170849.0$-$400910 & 17:08:47 & $-$40:08:51 & 51379.34 & 16800 & 0.47 & AXP J1708$-$4008 \\
SNR	 & G327.1$-$1.1            & 15:54:24 & $-$55:04:04 & 51382.22 & 16800 & 0.41 & \ldots \\
AXP	 & 1E1841$-$045            & 18:41:19 & $-$04:56:09 & 51382.44 & 16800 & 0.43 & AXP J1841$-$0456 \\  
SNR	 & G347.5$-$0.5$^{c}$      & 17:13:28 & $-$39:49:44 & 51383.29 & 16800 & 0.47 & \ldots \\
AXP$^{a}$ & AX J1845.0$-$0258  & 18:44:53 & $-$02:56:40 & 51391.45 & 16800 & 0.42 & \ldots 
\enddata
\tablecomments{All observations used a bandwidth of 288 MHz split into 96 3-MHz channels. A sampling time of 0.25 ms was used unless otherwise noted. The quoted fluence limits at 1374 MHz assume a pulse width of 1 ms and account for the different sky temperatures at the different target locations. \\
$^{b}$Also known as RX J0852.0$-$4622, \\
$^{a}$AX J1845.0$-$0258 is an AXP candidate, \\
$^{c}$A sampling time of 0.5 ms used.
}
\end{deluxetable}

\bibliography{c23}{}

\begin{thebibliography}{}
\expandafter\ifx\csname natexlab\endcsname\relax\def\natexlab#1{#1}\fi
\providecommand{\url}[1]{\href{#1}{#1}}
\providecommand{\dodoi}[1]{doi:~\href{http://doi.org/#1}{\nolinkurl{#1}}}
\providecommand{\doeprint}[1]{\href{http://ascl.net/#1}{\nolinkurl{http://ascl.net/#1}}}
\providecommand{\doarXiv}[1]{\href{https://arxiv.org/abs/#1}{\nolinkurl{https://arxiv.org/abs/#1}}}

\bibitem[{{Agarwal} {et~al.}(2020){Agarwal}, {Aggarwal}, {Burke-Spolaor},
  {Lorimer}, \& {Garver-Daniels}}]{aab+20}
{Agarwal}, D., {Aggarwal}, K., {Burke-Spolaor}, S., {Lorimer}, D.~R., \&
  {Garver-Daniels}, N. 2020, \mnras, 497, 1661, \dodoi{10.1093/mnras/staa1856}

\bibitem[{{Barsdell}(2012)}]{b12}
{Barsdell}, B.~R. 2012, PhD thesis, Swinburne University of Technology

\bibitem[{{Barsdell} {et~al.}(2012){Barsdell}, {Bailes}, {Barnes}, \&
  {Fluke}}]{bbb+12}
{Barsdell}, B.~R., {Bailes}, M., {Barnes}, D.~G., \& {Fluke}, C.~J. 2012,
  \mnras, 422, 379, \dodoi{10.1111/j.1365-2966.2012.20622.x}

\bibitem[{{Burgay} {et~al.}(2006){Burgay}, {Rea}, {Israel}, {Possenti},
  {Burderi}, {di Salvo}, {D'Amico}, \& {Stella}}]{bri+06}
{Burgay}, M., {Rea}, N., {Israel}, G.~L., {et~al.} 2006, \mnras, 372, 410,
  \dodoi{10.1111/j.1365-2966.2006.10872.x}

\bibitem[{{Camilo} {et~al.}(2009){Camilo}, {Ransom}, {Gaensler}, \&
  {Lorimer}}]{crg+09}
{Camilo}, F., {Ransom}, S.~M., {Gaensler}, B.~M., \& {Lorimer}, D.~R. 2009,
  \apjl, 700, L34, \dodoi{10.1088/0004-637X/700/1/L34}

\bibitem[{{Cordes} \& {Lazio}(2002)}]{cl02}
{Cordes}, J.~M., \& {Lazio}, T.~J.~W. 2002, arXiv e-prints, astro,
  \dodoi{10.48550/arXiv.astro-ph/0207156}

\bibitem[{{Crawford} {et~al.}(2007){Crawford}, {Hessels}, \& {Kaspi}}]{chk07}
{Crawford}, F., {Hessels}, J. W.~T., \& {Kaspi}, V.~M. 2007, \apj, 662, 1183,
  \dodoi{10.1086/517991}

\bibitem[{{Crawford} {et~al.}(2002){Crawford}, {Pivovaroff}, {Kaspi}, \&
  {Manchester}}]{cpk02}
{Crawford}, F., {Pivovaroff}, M.~J., {Kaspi}, V.~M., \& {Manchester}, R.~N.
  2002, in Astronomical Society of the Pacific Conference Series, Vol. 271,
  Neutron Stars in Supernova Remnants, ed. P.~O. {Slane} \& B.~M. {Gaensler},
  37.
\newblock \doarXiv{astro-ph/0111389}

\bibitem[{{Haslam} {et~al.}(1982){Haslam}, {Salter}, {Stoffel}, \&
  {Wilson}}]{hss+82}
{Haslam}, C.~G.~T., {Salter}, C.~J., {Stoffel}, H., \& {Wilson}, W.~E. 1982,
  \aaps, 47, 1

\bibitem[{{Lorimer} \& {Kramer}(2012)}]{lk12}
{Lorimer}, D.~R., \& {Kramer}, M. 2012, {Handbook of Pulsar Astronomy}
  (Cambridge University Press)

\bibitem[{{Manchester} {et~al.}(2001){Manchester}, {Lyne}, {Camilo}, {Bell},
  {Kaspi}, {D'Amico}, {McKay}, {Crawford}, {Stairs}, {Possenti}, {Kramer}, \&
  {Sheppard}}]{mlc+01}
{Manchester}, R.~N., {Lyne}, A.~G., {Camilo}, F., {et~al.} 2001, \mnras, 328,
  17, \dodoi{10.1046/j.1365-8711.2001.04751.x}

\bibitem[{{Staveley-Smith} {et~al.}(1996){Staveley-Smith}, {Wilson}, {Bird},
  {Disney}, {Ekers}, {Freeman}, {Haynes}, {Sinclair}, {Vaile}, {Webster}, \&
  {Wright}}]{swb+96}
{Staveley-Smith}, L., {Wilson}, W.~E., {Bird}, T.~S., {et~al.} 1996, \pasa, 13,
  243, \dodoi{10.1017/S1323358000020919}

\bibitem[{{Yao} {et~al.}(2017){Yao}, {Manchester}, \& {Wang}}]{ymw17}
{Yao}, J.~M., {Manchester}, R.~N., \& {Wang}, N. 2017, \apj, 835, 29,
  \dodoi{10.3847/1538-4357/835/1/29}

\end{thebibliography}
\bibliographystyle{aasjournal}
\end{document}